\documentclass[sn-mathphys,Numbered]{sn-jnl}
\usepackage{mathtools}
\usepackage{graphicx}
\usepackage{multirow}
\usepackage{amsmath,amssymb,amsfonts}
\usepackage{dirtytalk}
\usepackage{amsthm}
\usepackage{mathrsfs}
\usepackage[title]{appendix}
\usepackage{xcolor}
\usepackage{textcomp}
\usepackage{manyfoot}
\usepackage{booktabs}
\usepackage{algorithm}
\usepackage{algorithmicx}
\usepackage{algpseudocode}
\usepackage{listings}
\usepackage{lmodern}
\usepackage{anyfontsize}
\usepackage{wrapfig}
\usepackage{float}
\usepackage{subcaption}
\usepackage{caption}

\begin{document}

\title[Incentive Mechanism for Mobile Crowd Sensing with Assumed Bid Cost Reverse Auction]{Incentive Mechanism for Mobile Crowd Sensing with Assumed Bid Cost Reverse Auction}

\author*[1]{\fnm{Jowa} \sur{Yangchin}}\email{jowyangchen@gmail.com}

\author[1]{ \fnm{Ningrinla} \sur{Marchang}}\email{nm@nerist.ac.in}
\equalcont{These authors contributed equally to this work.}

\affil[1]{\orgdiv{Department of Computer Science \& Engineering}, \orgname{North Eastern Regional Institute of Science \& Technology}, \orgaddress{\street{Nirjuli}, \postcode{791109}, \state{Arunachal Pradesh}, \country{India}}}

\abstract{Mobile Crowd Sensing (MCS) is the mechanism wherein people can contribute in data collection process using their own mobile devices which have sensing capabilities. Incentives are rewards that individuals get in exchange for data they submit. Reverse Auction Bidding (RAB) is a framework that allows users to place bids for selling the data they collected. Task providers can select users to buy data from by looking at bids. Using the RAB framework, MCS system can be optimized for better user utility, task provider utility and platform utility. In this paper, we propose a novel approach called Reverse Auction with Assumed Bid Cost (RA-ABC) which allows users to place a bid in the system before collecting data. We opine that performing the tasks only after winning helps in reducing resource consumption instead of performing the tasks before bidding. User Return on Investment (ROI) is calculated with which they decide to further participate or not by either increasing or decreasing their bids. We also propose an extension of RA-ABC with dynamic recruitment (RA-ABCDR) in which we allow new users to join the system at any time during bidding rounds. Simulation results demonstrate that RA-ABC and RA-ABCDR outperform the widely used Tullock Optimal Prize Function, with RA-ABCDR achieving up to 54.6\% higher user retention and reducing auction cost by 22.2\%, thereby ensuring more efficient and sustainable system performance. Extensive simulations confirm that dynamic user recruitment significantly enhances performance across stability, fairness, and cost-efficiency metrics.}

\maketitle
\section{Introduction}\label{sec1}
\hspace{0.5cm} The rise in use of mobile devices has increased drastically in the last decade. Most new devices have many sensing capabilities which make them suitable for large scale data collection over a period of time. Community sensing refers to the process of sensing data of large-scale phenomena which cannot be sensed and collected by individual users. For example, the sensing of air pollution in an area can be done through community sensing. Community sensing involves an individual’s active contribution performing sensing tasks or autonomous sensing tasks where there is little to no active involvement by the individual categorized as participatory sensing [2] or opportunistic sensing [1]. The term MCS is used to refer to a broad range of community sensing paradigms [3]. The three main components of Mobile Crowdsensing are the Mobile users, the Task providers and the Sensing platform. Sensing platform is responsible for management of the system which involves receiving the sensing task request from the task providers, publishing those tasks to the mobile users, collection of the sensed data from the mobile users, responding the requesters with data and rewarding the mobile users for the task performed. As data collection is done by the mobile users who are individuals dispersed over a large area with different devices and various sensing capabilities, the rewards given to them as an incentive in exchange of their device power consumption and their own efforts and time is the only motivation for them to perform the tasks. The different mechanisms by which the users are given rewards are called the incentive mechanisms in mobile crowdsensing system. There has been a lot of work done to design different incentive mechanisms that better suit the system but there are many challenges yet to be solved. The main objective of designing incentive mechanisms is to collect more data with less incentives. Incentive mechanisms are classified into monetary and non-monetary incentives. Monetary incentive has a limit on budget and hence it is even more important to design efficient mechanisms. They are categorized into platform centric or user centric incentives. Platform centric incentive mechanisms aim to increase the platform utility while user centric mechanism focus more on increasing user profit and hence increase user utility. Both platform and user centric incentive mechanisms can be either dynamic, i.e., changing incentives or static, i.e., fixed incentives. The data collected must also be valid and of good quality. But while designing incentive mechanisms, generally, the focus is not on truthfulness of the users but on budget management.  

Auction theory is largely studied for the purpose of designing incentive mechanisms in mobile crowd sensing. A general auction system has a seller who possesses the item and buyers who place bids for the item. The item is sold to the buyer with highest bid. In a reverse auction system, the sellers have the items, and they place the bids for the item. The buyer selects the lowest bid to buy the item. 

Reverse auction is deployed as another approach towards designing incentive mechanisms using auction theory. Reverse auction was first used in participatory sensing in [15] as a way to have auctions in a large scale data collection process where the sellers are in control of the item price, which in this case is the sensed data cost. Reverse auction in MCS systems helps select the participants from whom the data will be collected. This system is used when there are many participants who are willing to submit the requested data. The mobile users are the sellers who have the sensed data as the item to be sold to the platform. They place a bid value for their data to the platform from which then the platform selects the item with lowest bid value and buys data. 
Moving away from the reverse auction mechanism, we propose a novel approach called Reverse Auction with Assumed Bid Cost (RA-ABC) which allows the users to place a bid in the system before performing the data collection task.

The key contributions of this paper can be summarized as:
\begin{enumerate}
    \item We propose RA-ABC, a novel incentive mechanism introducing a pre-bidding approach, allowing participants to submit bids before performing sensing tasks, unlike traditional reverse auctions.
    \item Our model incorporates bid adjustment strategies to optimize auction cost, enhance user participation, and balance engagement with efficiency.
    \item We extend RA-ABC to RA-ABCDR, integrating a dynamic recruitment mechanism that enables new participants to join in later auction rounds, ensuring system stability.
    \item The dynamic recruitment model improves long-term fairness by mitigating user drop-off, preventing monopolization, and fostering equitable auction participation.
    \item RA-ABCDR optimizes cost efficiency, achieving higher utility at lower auction costs than RA-ABC, making it more sustainable for extended deployments.
\end{enumerate}

\section{Related Work}\label{sec2}
\hspace{0.5cm} An MCS system with monetary incentive is limited by the budget available for the task. Many researchers have developed auction based incentive mechanisms for the purpose of improving budget efficiency. 
Random Selection based Fixed Pricing, RSFP \cite{b19} is a mechanism where a fixed amount incentive is paid to randomly selected required number of participants. This mechanism is fast but due to fixed reward, some users who have higher expectations might lose interest to participate further and drop out. There might be users who have lower expectations but who never get selected as the selection is random and hence, they too drop out. These limitations are removed by the method of Reverse Auction with Dynamic Pricing (RADP) \cite{b15} where the selection of winners for each round depending on the Bid value placed by participants at the beginning of auction round. After each round, participants are allowed to change their bids to increase their chances of winning or to increase their earned profit. While in this approach, there is no random selection of winners but still some users might leave the system on continuously losing the auction rounds. To keep such users from leaving the system, RADP with virtual participation credit (RADP-VPC) \cite{b15} is developed in which a fixed virtual credit is given to the losers in each auction round. The virtual credit does not represent a true reward but it can be used to decrease their bid for the next rounds and hence increasing their chances of winning. The users now place a virtual bid, which is equal to their virtual credit subtracted from their actual bid, this decreases their bid values. If a participant does not win for consecutive rounds, their virtual credit increases and decreases the virtual bid and hence increases their winning chance. Once they win, the virtual credit is reset to null value. This mechanism helps in retaining the users in the system for further participation. After each round, participants calculate their return on investment (ROI) \cite{b20} value using the total earned credit till that round and if they find that it is below their satisfaction, they choose to drop out of the system. To make the participants who decide to drop out after a round, to rejoin the system, the least bid, that is the top winning bid is revealed to them. Now these participants again calculate their ROI value using the revealed bid. This extension in RADP-VPC helps in recruiting the participants who drop out. With this information participants can now manipulate their own bid values by bidding a value close to the last winning bid such that it does not fall below their own true value of the data and choose to stay in the system with increase in their probability of winning. 

With RADP-VPC, the selection of winners is not random anymore and retaining and recruiting methods use the ROI calculation, but they do not consider each the participant’s contribution while making the choices. Hence, reverse auction based incentive mechanism (RAIN) [9] uses the participants' potential contribution while recruiting participants, selecting winners and retaining participants. It measures the impact of participants joining and staying in the system based on their inadequacy of task execution in the Positions of Interest (PoI), i.e., the place of their tasks. 

Reverse auction gives the power of data cost control to mobile users rather than the platform which attracts more users to start participating in the system. The use of earned profit till current round for calculation of ROI value in RADP-VPC and RAIN needs the system to keep track of total earnings for all the participants. This approach becomes inconvenient when the number of rounds and the number of participants is very large. Also, recruiting participants based on the user’s rationality results in weak platform utility. So, determining a recruitment process that does not impact the budget balance is a problem. Retaining existing members who know the task so that the workers composition can be optimized is another challenge. Existing works do not consider participants’ history of participation or winning in past auction rounds which is a valuable information on user’s behavior in context with their performance. Also, in the existing systems, although there is mechanism for retaining and recruiting users, the participant pool is fixed, from where they can recruit the drop out users. A MCS system with reverse auction should allow new users to join or leave the system whenever they wish as all users are not in the same POI at all times. Using reverse auction in MCS requires the participants to know the true value of the task so that they can place a bid, so the existing systems assume that users who come to participate already have done the task and know their true value. But in the real world, many users may not be willing to perform the tasks if they are not certain that they can earn something for it. 

To address the aforementioned challenges, we propose an incentive mechanism called Reverse Auction- Assumed bid cost (RA-ABC) wherein the users do not need to perform the task before bidding. Having their device capability information and their own potential POIs, the users are allowed to bid in the system with a bid value that they estimate with an assumed true value given the task information. Once they win a round, participants who won can perform the task with surety of rewards. For this system though, for avoiding unnecessary bids from malicious users a participation cost is charged. The value of participation cost is insignificant compared to the reward value, so that users are still encouraged to place the bids. We use participation history and average profit of users to calculate the ROI value at each round. This ensures that those users who have participated more frequently, thus having better potential to contribute, are favored to stay in the system.

\section{RA-ABC (Reverse Auction - Assumed Bid Cost)}\label{sec3}
\hspace{0.5cm} This section elaborates on RA-ABC system model: assumed bid cost (ABC), auction round winner selection and participant retaining.
The task provider first sends the task to the sensing platform, which publishes the task to all mobile users. Users who choose to participate in the sensing task then place a bid to the system. The auction begins when the sensing platform receives the bids and selects the winner bids. Winners are rewarded the incentive after they submit the sensed data. This completes one round of auction in the system. The system runs for multiple rounds where each round selects different winners depending on their bids.

\subsection{Assumed Bid Cost (ABC) }\label{subsec3}
\hspace{0.5cm} Generally, in a reverse auction bidding system, users already possess the data they wish to sell and know its true cost, \emph{(c)}, placing bids accordingly to earn a reasonable profit. However, a major drawback in such a system is that users must collect data before participating in the auction, meaning their effort is unrewarded if they do not win. For example, in a participatory sensing environment, a user performs a task and submits data before the deadline, but if they lose the auction, their work goes to waste. 
To address this inefficiency, users in RA-ABC are allowed to place bids before collecting data, estimating an assumed true cost,$(\bar{c})$, and calculating an approximate bid cost accordingly. The value of $\bar{c}$ is derived from historical bid estimates. 
To ensure that assumed bid costs accurately reflect actual data collection costs, an adaptive cost adjustment mechanism with a bid revision window is implemented. Initially, users submit bids based on estimated factors such as resource consumption, travel expenses, and task complexity, using historical cost data for refinement. 
Each participant provides an initial bid $(b_i^{(0)})$, calculated via:
\begin{equation}
    b_i^{r(0)} = E[b] + \alpha \cdot V[b] \label{eqn2}   
\end{equation}
where \emph{E[b]} represents historical bid estimates from past auctions,
\emph{V[b]} accounts for cost variability due to task uncertainties,
$\alpha$ adjusts for risk tolerance, with higher values leading to conservative bidding.
Before the final bid submission deadline, users enter a revision phase, where they can update bids based on real-time observations such as network latency or battery constraints. Bid refinement follows a Bayesian update model, adjusting values dynamically as new observations, \emph{x}, become available.
\begin{equation}
    b_i^{r} = b_i^{(r-1)} + \frac{\sigma^2}{\sigma^2 + \tau^2}(x-b_i^{r-1})\label{eqn2-1}
\end{equation}
where $\sigma^2$ is the variance of prior bid estimates,
$\tau^2 $ represents variance in newly observed environmental data.

Strategic bid manipulation prevention is further reinforced using deviation penalties:
\begin{equation}
    p_i = \gamma \cdot |b_i^{r} - b_i^{r(0)}| \label{eqn2-2}   
\end{equation}
where:
$p_i$ imposes ranking penalties for excessive bid changes,
$\gamma$ scales sensitivity to bid adjustments.
Finally, probabilistic cost estimation models, such as Bayesian updates or truncated normal distributions, refine sensing cost predictions dynamically, giving preferential rankings to users with historically accurate bids. After the review window closes, the final bids are submitted, and the auction mechanism selects the the winners, ensuring fairness and cost accuracy. Ultimately, the adjusted bid $(b_i^{r})$ is finalized, promoting optimized pricing and rational bidding behavior throughout the auction process. 
To quantify how bid adjustments influence auction outcomes, we define the bid adjustment impact (BAI) as:
\begin{equation}
    BAI = \frac{(b_i^r - b_i^{r(0)})}{b_i^{r(0)}} \times P(w) \label{eqn2-22}
\end{equation}
where $ b_i^{r(0)}$ represents the user’s initial bid before any modifications.
$b_i^r$ denotes the adjusted bid after revision.
$P(w)$ is the winning probability, derived from historical bid success rates.
$P(w) $ serves as metric to track how bid adjustments influence auction success. This formulation enables a systematic evaluation of how bid revisions affect strategic positioning and success within the auction process.
We now present the following theorem, which establishes fundamental properties of the adaptive bid cost function in the bid-before-acting mechanism:

$\mathbf{Theorem1}$ The adaptive bid cost function $b_i^{r}$ in the bid-before-acting mechanism is a mono, non-negative and submodular function.

\emph{Proof: Non-negativity}
Since the bid cost estimation follows:
\begin{equation}
    b_i^{r(0)} = E[C] + \alpha V[C]\label{eqn3}
\end{equation}
where:
$E[C]$ is the expected historical cost (always nonnegative),
$V[C]$ is the variance (nonnegative),
$\alpha$ is a risk factor (nonnegative),
Thus, $b_i^{r(0)} \geq 0$. Moreover, in each bid revision window, the Bayesian update ensures:
\begin{equation}
    b_i^{r} = b_i^{(r-1)} + \frac{\sigma^2}{\sigma^2 + \tau^2} (x - b_i^{(r-1)})\label{eqn4}
\end{equation}
where $\sigma^2, \tau^2 \geq 0$. Since \emph{x} represents observed environmental feedback, the bid cost remains nonnegative for all \emph{t}:
$b_i^{r} \geq 0 \quad \forall r.$

\emph{Proof: Monotonicity}
For all bid rounds \emph{t}, Bayesian updates refine cost estimation:
$b_i^{r} \geq b_i^{(r-1)}$
because the correction term $\frac{\sigma^2}{\sigma^2 + \tau^2} (x - b_i^{(r-1)})$ adjusts bids only if real-time observations \emph{x} suggest an increase. This ensures monotonic increase, preventing bid reductions:
$b_i^{r} \geq b_i^{(r-1)}, \quad \forall r.$

\emph{Proof: Sub modularity}
For participants $i$ and $j$, let:
$b_{i,j}^{r} = b_i^{r} + b_j^{r}.$
Since bids are updated adaptively via Bayesian inference, adding more participants does not increase marginal returns linearly. Instead, it follows diminishing returns:
$b_{i,j}^{r} - b_i^{r} \leq b_j^{r}.$
Thus, the incremental bid cost addition is submodular:
$b_{i,j}^{r} \leq b_i^{r} + b_j^{r}.$
Hence, the adaptive bid function exhibits monotonicity, nonnegativity, and submodularity, ensuring fairness in long-term auctions.

\subsection{Winner Selection}\label{subsec4}
\hspace{0.5cm} The selection process is influenced by two key factors: (a) the bid value and (b) the user's history of participation in previous auction rounds. Users with a strong participation record are given preference over those who are infrequent participants, ensuring fair competition and sustained engagement. The next subsection details how participation frequency is calculated using the Exponentially Weighted Moving Average (EWMA) method \cite{b17}.
The EWMA method is employed in the RA-ABC system to evaluate participant frequency and Return on Investment (ROI) due to its adaptability, responsiveness, and ability to smooth fluctuations effectively. It applies historical weighting with exponential decay, ensuring that recent participation events have greater influence than older ones, allowing the system to track active users accurately while mitigating outdated biases. 
Let $e_r$ represent the event at round \emph{r}. Thus $e_r$=1 if user participate in round \emph{r}; $e_r$=0 otherwise. EWMA computes a weighted average of the sequence by applying weights that decrease geometrically with the age of the observations \cite{b17}. If a user has participated in recent rounds, her EWMA of participation frequency is higher than of those who have participated the same number of times in the auction but not recently. Thus, the average participation frequency of user \emph{i} at round \emph{r}, $p_i^r$ is defined as:
\begin{equation}
     p_i^r= \alpha . e_r + (1 - \alpha) . p_i^{r-1} \label{eq5}
\end{equation}
\noindent where $\alpha$ is the weighting factor having a value between 0 and 1. The closer $\alpha$ is to 1, the more weight is given to recent data points. Here, \emph{$e^r$} represents the event of participation, having the value of 0 or 1 at round \emph{r} and \emph{$ p_i^{r-1}$} is the participation frequency at round \emph{$r-1$}. We initialize $p_i^0$=0.
This approach prevents sudden dropouts by maintaining a fair participation tracking mechanism, ensuring that users are not penalized for temporary inactivity as a pure arithmetic average would. Additionally, EWMA stabilizes ROI computation, ensuring gradual adjustments to incentive structures and preventing abrupt rank declines due to short-term fluctuations.
Polynomial-based methods, such as quadratic regression or higher-order trend analysis, were avoided due to their susceptibility to overfitting, which causes participation trends to become rigid and fail to adapt to real-time auction dynamics. Moreover, polynomial methods introduce significant computational overhead, making real-time auction adjustments impractical. Additionally, EWMA mitigates bid manipulation risks within the bid revision window, ensuring consistent tracking. Unlike EWMA, polynomial models lack exponential decay, limiting their ability to dynamically adjust participation history.
Ultimately, EWMA enhances fairness, improves responsiveness, and provides a stable framework for evaluating auction participants.

\subsection{Retaining Users}\label{subsec5}
\hspace{0.5cm} If a user's bid consistently fails to secure a winning position, even with a high tolerance for losses, they will eventually decide to leave the system. If multiple users exit after only a few rounds, the auction process cannot sustain itself over an extended duration. To ensure long-term engagement, there must be a mechanism that allows users to assess their continued participation. The platform achieves this by calculating the Return on Investment (ROI), which serves as a measure of user satisfaction. Its represents utility of a user. The ROI quantifies the profit derived from investment \cite{b20}, helping users determine whether remaining in the system is beneficial. The following subsections outline the formulation of ROI in RA-ABC.
\subsection{Average Earnings}
\hspace{0.5cm} The ROI computation is based on the average earnings of a user over multiple rounds, representing long-term profitability. To capture recent performance trends, the Exponentially Weighted Moving Average (EWMA) of winning bids is employed, as users primarily evaluate their earnings from recent auction rounds when determining system satisfaction.
The ROI calculation in RA-ABC emphasizes average earnings and participant frequency, deliberately excluding historical cost summation to uphold fairness and encourage long-term engagement. Summing past costs could disproportionately penalize users who previously incurred high expenses but later adapted their bidding strategies, limiting their ability to remain competitive. Instead, the EWMA-based approach ensures that recent performance has a greater impact on ROI calculations, preventing outdated expenditures from distorting assessments.
Additionally, stability in user incentives and retention is reinforced by avoiding cumulative cost tracking, which could lead to declining ROI values for long-term participants, discouraging sustained engagement. The EWMA method smooths fluctuations, preventing participants from being unfairly penalized for temporary inefficiencies.
The adaptive cost impact mitigation accounts for real-world sensing challenges, such as energy depletion and environmental variability, which cause sharp cost fluctuations. Rather than relying on an absolute cost summation, the system adjusts ROI dynamically based on expected earnings and participation frequency, mitigating cost-related biases.
Furthermore, the approach enhances auction stability, as direct historical cost inclusion may lead to strategic bidding manipulation, where users initially underestimate costs to remain competitive before inflating them later. By prioritizing consistency and fairness, RA-ABC fosters a transparent, balanced bidding environment, ensuring that users compete based on strategy and efficiency rather than exploitative cost manipulation. 

To calculate the average earnings, $\emph{m}_i^r$ for each user \emph{i} at round \emph{r}, we use:
\begin{equation}
 m_i^r= 
\begin{dcases}
     \beta . b_i^{r} + (1 - \beta) . m_i^{r-1} , &\text{if user } i \text{ wins in round } r\\
      (1 - \beta) . m_i^{r-1}     &\text{otherwise}
\end{dcases}
\label{eq7}
\end{equation}
\noindent where $b_i^r$ is the bid placed by user \emph{i} at round \emph{r}. Here, $\beta$ is the weighing factor. 

The calculation of ROI value for a user helps them decide if they can earn a satisfactory reward after they bid on the next round even if they have incurred a loss in the current round. This value depends on the ratio of average of bids placed by the user till that round to what they expected to earn. Their expected earnings are represented by the frequency of participation and their actual true value. A satisfaction threshold, S, is set by the system below which if the ROI falls, then users leave the system as they are not satisfied by the system. The higher the satisfaction threshold, the lower will be the number of participants as lesser participants will be satisfied. When a user joins the system, their ROI value is set to a value that is slightly higher than the threshold value so that they do not leave the system immediately after loosing the first round. To select a satisfaction threshold for the system is a challenge as not all users have the mobile devices with same capabilities. If the satisfaction threshold is set higher, as the number of participants decreases, the auction cost for the rounds increases. This happens because lower number of participants results in lesser competition between the users. Then the probability of winning increases and the users still win with higher bid. 

\subsubsection{ROI Calculation by Active Participants}\label{ROI}

\hspace{0.5cm} Assume an user who is actively participating in the auction. In other words, she is bidding and thus seeking a chance to win. We call such participant as an active participant. Let $\eta_i^r$  denote the ROI for a user \emph{i} in round \emph{r} which is defined as: 
\begin{equation}
  \eta_i^r = \frac{m_i^r + \tau_i}{p_i^r. c_i + \tau_i} \label{eq3} 
\end{equation}

\noindent where $\emph{m}_i^r$ and $p_i^r$ represent weighted average earnings and frequency of participation of user \emph{i} in round \emph{r} respectively (please refer Eq. \ref{eq5} and Eq. \ref{eq7} for their formulas). Here $c_i$ represents the true cost of the task performed by user \emph{i}. This value is known only after she performs the task. Moreover $\tau_i$ represents the tolerance value of user \emph{i}. This value gives a measure of how much a user is willing to tolerate losing a bid. For a user with higher value of $\tau$, the ROI value will decrease slower. When the ROI value drops below a pre-defined satisfaction threshold (S), the user drops out of the system. The value of S is set to a constant value for all users.

\subsubsection{ROI Calculation by Dropped Participants}
\hspace{0.5cm} When an active participant's ROI value goes below the threshold (S), it drops out. Thus it becomes a dropped participant. However it may wish to rejoin the auction.Thus, a dropped user may not be participating in current round \emph{r}. However, to decide whether to rejoin, it estimates the ROI for the next round \emph{r+1}. If the expected ROI value is not less than the threshold value, the user rejoins the system and participates in the next round. To calculate the ROI estimate $\eta_i^{r+1}$ of user \emph{i} in round \emph{r+1} we use:
\begin{equation}
  \eta_i^{r+1} = \frac{m_i^{r+1} + \tau_i}{p_i^{r+1}. \bar{c_i} + \tau_i} \label{eq8} 
\end{equation}
where $m_i^{r+1}$ and $p_i^{r+1}$ are the weighted average earning and frequency of participation of user \emph{i} in round \emph{r+1} respectively. Here $\bar{c_i}$ is the estimate of the true cost of the task. Since user \emph{i} has not performed the task yet, it uses an estimate of the true value. The tolerance value of \emph{i} is denoted by $\tau_i$.

\section{Dynamic Participant Recruitment}\label{sec4}
\hspace{0.5cm} RA-ABCDR (Reverse Auction—Assumed Bid Cost with Dynamic Recruitment) enhances competition, fairness, and system stability in Mobile Crowd Sensing (MCS) by introducing dynamic user recruitment. Unlike RA-ABC, where the participant pool remains fixed after initialization, RA-ABCDR allows users to join at any point, submit bids, and have their Return on Investment (ROI) calculated in the next round. If their ROI is unsatisfactory, they may exit. This mechanism prevents monopolization, ensures a fluid marketplace, and balances participant numbers dynamically over time. By continuously integrating new users, RA-ABCDR maintains contributor diversity, prevents stagnation, and sustains auction activity and sensing participation. Additionally, adaptive auction scaling keeps sensing capacity flexible, preventing user shortages during critical periods.
\subsection{System Stability}
\hspace{0.5cm} The RA-ABCDR auction framework enhances long-term resilience through dynamic recruitment, fostering participant diversity and preventing stagnation. By continuously introducing new contributors, the system avoids dependency on a fixed user group, ensuring varied sensing capabilities and competitive engagement. Lower dropout rates help sustain auction activity, as discouraged users can be seamlessly replaced with fresh bidders. Additionally, adaptive auction scaling strengthens system stability by maintaining flexible sensing capacity, preventing resource shortages during critical sensing phases.
Two key indicators of system stability in Mobile Crowd Sensing (MCS) are the total cost of tasks called auction cost and the number of users participating in each auction round. These metrics help evaluate how efficiently the auction mechanism allocates sensing resources and whether user engagement remains consistent over time. A system with stable costs and active participation is more likely to uphold fairness, efficiency, and long-term sustainability.
The auction cost represents the cumulative sensing expenses incurred across all active participants in a given round, as defined in Eq.[\ref{eqn1-1}]. 
\begin{equation}
    C_{auc}=\sum b_i \quad \forall\quad  i \in winner \label{eqn1-1}
\end{equation}
The number of users per round serves as a key measure of system engagement and accessibility. Without mechanisms to broaden participation, fewer users may control the market, undermining fairness. To address this, adaptive dynamic recruitment strategy is used to attract new users while balancing incentives for existing participants.

\subsection{Fairness}
\hspace{0.5cm} Dynamic recruitment enhances auction fairness by ensuring equal bidding opportunities for new entrants, preventing monopolization by experienced users. Without this mechanism, high-frequency participants could develop dominant bidding strategies, undermining competition. Additionally, new users disrupt strategic bidding manipulation, reducing collusion risks and fostering a more transparent, competitive environment. By maintaining a steady influx of diverse bidders, RA-ABCDR ensures that auctions remain fair, with balanced opportunities for both new and established users.

We use the Monopoly Prevention Index (MPI) defined in Eq.(\ref{eqn8}) to assess fairness in the auction-based systems. MPI quantifies how equitably wins are distributed among participants, ensuring that auctions remain competitive and no single user disproportionately dominates over multiple rounds. By analyzing win frequencies, MPI helps maintain balanced participation, fostering long-term engagement and preventing monopolization within the system. The formulation of MPI draws inspiration from concentration indices like the Herfindahl-Hirschman Index (HHI) \cite{b22} but extends it by incorporating an explicit penalty on the most dominant participant, thereby addressing both distributional inequality and excessive winner bias. As a result, MPI provides a more comprehensive fairness metric tailored for dynamic, multi-round auctions in mobile crowd sensing systems. A higher MPI indicates a fairer distribution of opportunities, while lower values reveal the risk of monopolistic behavior.
\begin{equation}
    MPI = 1 - \frac{\sum_{i=1}^N w_i^2}{N} - \left| \frac{\max(w_i)}{\sum_{i=1}^{N} w_i} - \frac{1}{N} \right| \label{eqn8}
\end{equation}
where $w_i$ is the win frequency of a user \emph{i} and $N$ is the total number of users in a given round. $1 - \frac{\sum_{i=1}^N w_i^2}{N}$ evaluates win distribution fairness by computing the squared sum of win frequencies across participants. If wins are evenly spread, the summation $( \sum w_i^2 )$ remains low, meaning MPI stays high, indicating fairness. If one or a few users dominate, $( \sum w_i^2 )$ increases, making MPI drop, signaling potential monopolization in the system.
This term alone provides a general fairness measure, but it does not specifically detect when a single participant excessively dominates. Therefore, $\left| \frac{\max(w_i)}{\sum_{i=1}^{N} w_i} - \frac{1}{N} \right|$ directly penalizes monopolization by adjusting MPI based on how much the top winner contributes to the total wins. The fraction $( \frac{\max(w_i)}{\sum w_i} )$ measures dominance—it checks the proportion of wins controlled by the top-performing user. The term $( \frac{1}{N} )$ represents ideal fairness, where all users win equally across rounds.
The difference between these two terms tells us how far reality deviates from fairness. The greater this difference, the lower MPI becomes, ensuring that monopolization is penalized more effectively.
We use Bid Accuracy Ratio (BAR) defined in Eq.(\ref{eqn9}) as another fairness metric in the system. It evaluates how closely users’ bids align with their actual sensing costs, ensuring rational bidding behavior while discouraging strategic price manipulation.
\begin{equation}
 BAR_i = \frac{|b_i - c_i|}{c_i}   \label{eqn9}
\end{equation}
 where $b_i$ is the user’s bid in a given round and $c_i$is the actual cost incurred for sensing. A lower BAR value indicates fair bidding, whereas a higher BAR suggests bid distortion, potentially signaling manipulation, or inefficiency.

\section{PERFORMANCE EVALUATION}\label{PE}
\hspace{0.5cm}In this section, we elaborate the experimental set up and results of performance evaluation. We compare the two proposed approaches:  (a) RA-ABC and (b) RA-ABCDR with the existing mechanism (c) tullock optimal prize function \cite{b21} for different performance metrics.

\subsection{Experimental Set-up}
\hspace{0.5cm} In order to verify the working of the proposed mechanisms, we carry out extensive simulation. All simulations and graph plots in this section are implemented using Python with Anaconda Navigator GUI using Spyder 5.1.5 IDE and executed on an AMD RYZEN5 PC with 8 GB RAM.

The simulation is set up using 100 initial bidders (\emph{N}) with required number of winners (\emph{W}) as 20. The auction is run for 100 rounds. For the initial round, assumed true cost, $\bar{c}$ for users are generated using Gaussian distribution with mean value 5. This value is for each user is between 90\% to 105\% of actual true cost \emph{c}. ROI threshold, \emph{S} is set to 0.5.

\subsection{Experimental Results}\label{ex}
\hspace{0.5cm} We simulate for 50 different scenarios. Thus each point in in a plot represents average of 50 runs. The analysis is done for key evaluation metrics: (a) System efficiency: measured by the user utility (refer section \ref{ROI} Eq. [\ref{eq3}]) and bid adjustment impact (BAI) from Eq. [\ref{eqn2-22}]. (b) System stability: the system stability is anlyzed by number of active participants in the system and total auction cost (refer Eq. [\ref{eqn1-1}]) in every round. (c) System fairness: the monopoly prevention index (MPI) from Eq.[\ref{eqn8}] and bid accuracy ratio (BAR) in Eq. [\ref{eqn9}] are used to analyze the fairness of the system. 

\subsubsection{System Efficiency and Stability}
\begin{figure}[htbp]
\begin{subfigure}[b]{0.5\textwidth}
    \centerline{\includegraphics[width=1\textwidth]{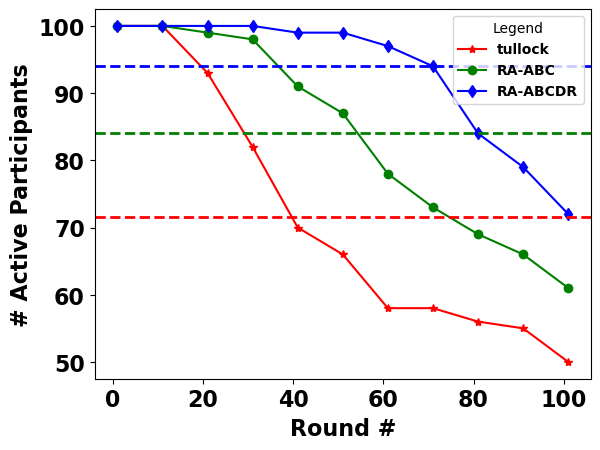}}
\caption{No. of Active Participants}
\label{fig1a}
\end{subfigure}
\begin{subfigure}[b]{0.5\textwidth}
    \centerline{\includegraphics[width=1\textwidth]{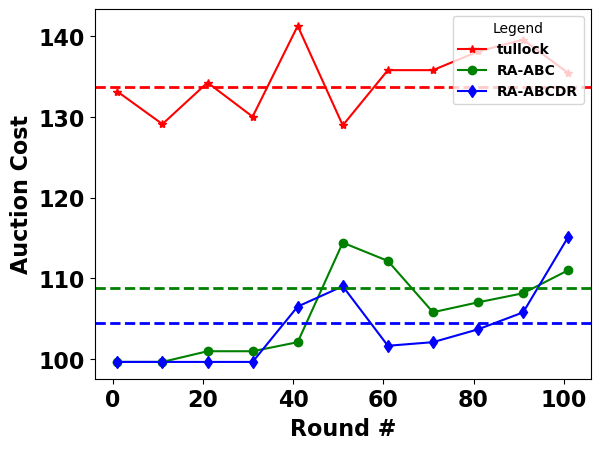}}
\caption{Auction Cost} 
\label{fig1b}
\end{subfigure}
\caption{System Stability: Comparison across different mechanisms}\label{fig1}
\end{figure}

\hspace{0.5cm} Fig.~\ref{fig1a} shows the number of active participants in different methods for 100 rounds of auctions. The graph shows the percentage difference in active participants between RA-ABC, RA-ABCDR, and Tullock relative to the average participation level (dashed line). RA-ABCDR retains 23.3\% more users than the average, demonstrating strong participant engagement and sustainability. RA-ABC remains close to the average with a 2.7\% increase, indicating moderate retention. In contrast, Tullock falls 24.7\% below the average, reflecting significant user drop-off over time. These differences highlight RA-ABCDR's superior ability to maintain user engagement, while RA-ABC provides moderate stability, and Tullock struggles with retention, making it the least effective in sustaining long-term participation. Fig. ~\ref{fig1b} illustrates that Tullock maintains the highest auction costs and RA-ABCDR achieves the lowest cost. The dashed lines mark the average auction cost for each method, highlighting that RA-ABCDR consistently optimizes bidding strategies, minimizing cost escalation compared to RA-ABC and Tullock. 

\begin{figure}[htbp]
\begin{subfigure}[b]{0.5\textwidth}
    \centerline{\includegraphics[width=1\textwidth]{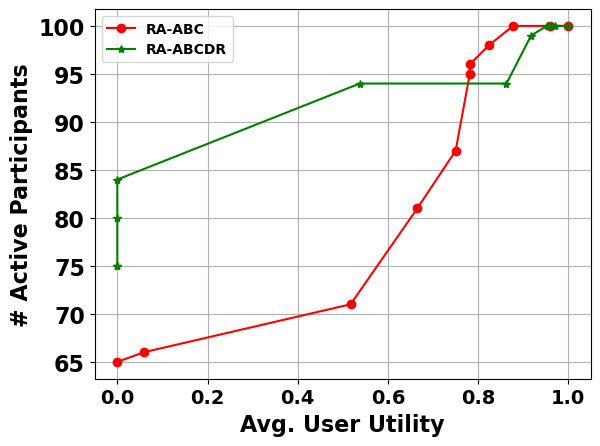}}
\caption{ User Utility }
\label{fig2a}
\end{subfigure}
\begin{subfigure}[b]{0.5\textwidth}
    \centerline{\includegraphics[width=1\textwidth]{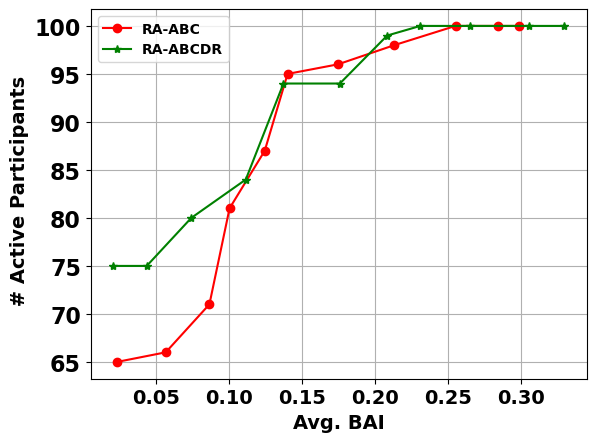}}
\caption{Bid Adjustment Impact (BAI)} 
\label{fig2b}
\end{subfigure}
\caption{System efficiency: Active Participants}\label{fig2}
\end{figure}

The graph in Fig.~\ref{fig2a} shows that as user utility increases, participation also rises, with RA-ABCDR consistently maintaining higher engagement than RA-ABC. RA-ABCDR's dynamic recruitment strategy prevents stagnation and ensures fairer auctions, while RA-ABC, though effective in bid optimization, struggles with long-term retention. This makes RA-ABCDR the better choice for adaptive engagement and auction stability. Fig.~\ref{fig2b} illustrates the relationship between average BAI and the number of active participants for RA-ABC and RA-ABCDR. As the average BAI increases, the number of participants rises for both mechanisms, indicating that bid adjustments contribute to higher engagement. However, RA-ABCDR reaches 100 active participants at a lower BAI value (~0.20) compared to RA-ABC (~0.25), suggesting that RA-ABCDR achieves effective bid optimization earlier, ensuring participation remains high even with lower bid adjustments.

\begin{figure}[htbp]
\begin{subfigure}[b]{0.5\textwidth}
    \centerline{\includegraphics[width=1\textwidth]{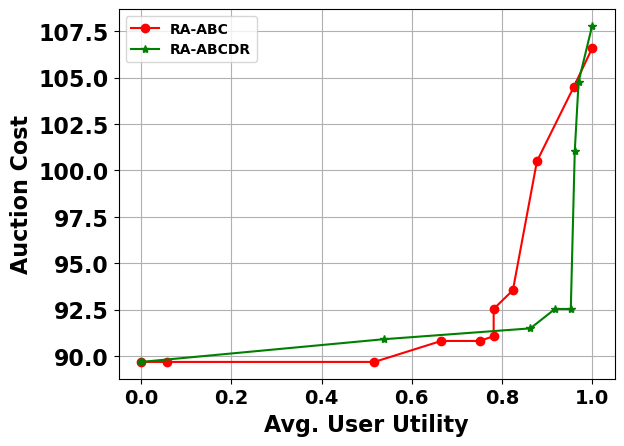}}
\caption{ User Utility }
\label{fig3a}
\end{subfigure}
\begin{subfigure}[b]{0.5\textwidth}
    \centerline{\includegraphics[width=1\textwidth]{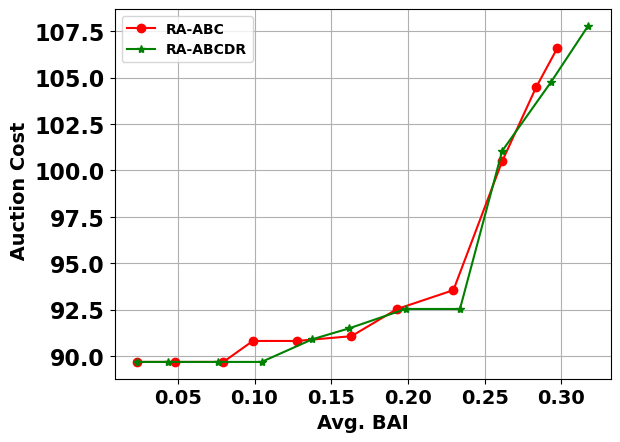}}
\caption{Bid Adjustment Impact (BAI)} 
\label{fig3b}
\end{subfigure}
\caption{System efficiency: Auction Cost}\label{fig3}
\end{figure}
 The graph in Fig.~\ref{fig3a} shows that RA-ABC leads to steeper auction cost increases, while RA-ABCDR maintains better cost efficiency as user utility rises. This suggests that RA-ABCDR optimizes bidding strategies, preventing excessive inflation while sustaining fairness and engagement, making it the better choice for long-term stability. Fig.~\ref{fig3b} shows that in RA-ABC, auction cost continues to rise even at lower utility levels, indicating less efficient cost scaling. In contrast, RA-ABCDR reaches a near-maximum utility (~0.98) at an auction cost of approximately 92.5, whereas RA-ABC requires a significantly higher cost to achieve the same utility. This demonstrates RA-ABCDR's superior cost-efficiency, ensuring users attain high utility without excessive bidding expenses.

\begin{figure}[htbp]
\begin{subfigure}[b]{0.5\textwidth}
    \centerline{\includegraphics[width=1\textwidth]{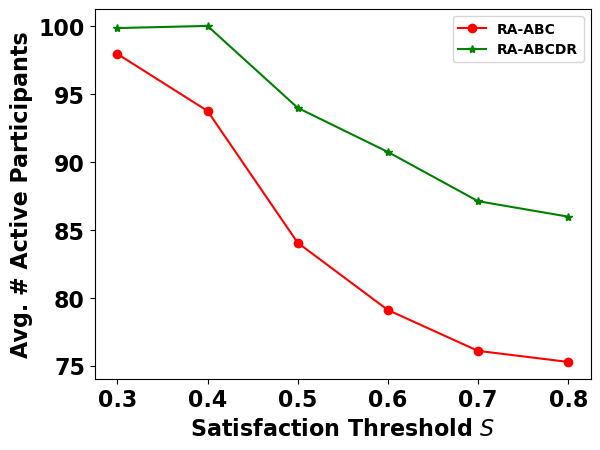}}
\caption{ No. of participants for varying S values}
\label{fig4a}
\end{subfigure}
\begin{subfigure}[b]{0.5\textwidth}
    \centerline{\includegraphics[width=1\textwidth]{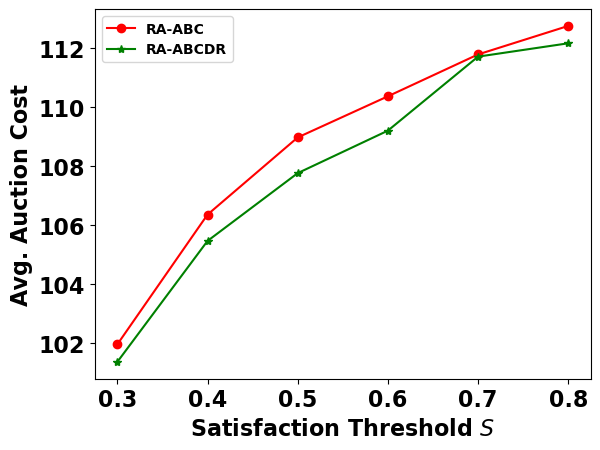}}
\caption{Auction cost for varying S values} 
\label{fig4b}
\end{subfigure}
\caption{Varying satisfaction threshold}\label{fig4}
\end{figure}

\paragraph{Effect of Satisfaction Threshold}\
The ROI threshold represents the measure of satisfaction of the participants. For higher values of S there will be less users who would stay in the system. As S increases, participant numbers decline for both methods as shown in Fig.~\ref{fig4a}, but RA-ABCDR retains more users at higher thresholds than RA-ABC. RA-ABC drops sharply from 100 to 75, while RA-ABCDR maintains greater engagement, decreasing to 90 at S = 0.8. This shows RA-ABCDR fosters better long-term retention, ensuring users remain active as satisfaction requirements rise. Its 20\% higher user retention highlights stronger adaptability and fairness in sustaining participation.
In Fig.~\ref{fig4b} as S increases, Auction Cost rises for both methods, but RA-ABC consistently leads to higher costs at every threshold level. This indicates that RA-ABCDR employs more efficient bidding strategies, keeping Auction Costs lower while still supporting increased participant satisfaction.

\subsubsection{System Fairness}

\begin{figure}[htbp]
\begin{subfigure}[b]{0.5\textwidth}
    \centerline{\includegraphics[width=1\textwidth]{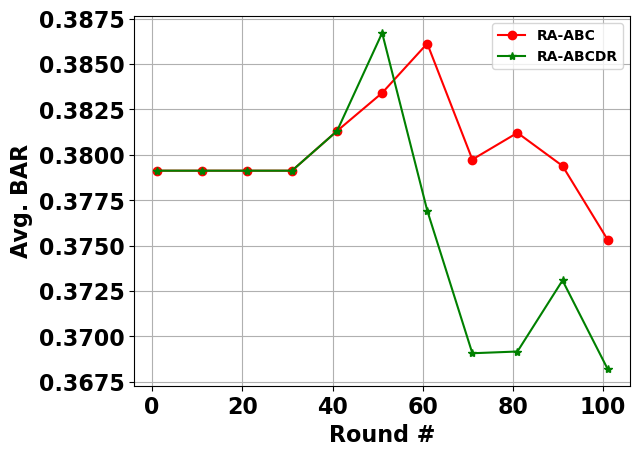}}
\caption{ Bid Accuracy Ration (BAR) }
\label{fig5a}
\end{subfigure}
\begin{subfigure}[b]{0.5\textwidth}
    \centerline{\includegraphics[width=1\textwidth]{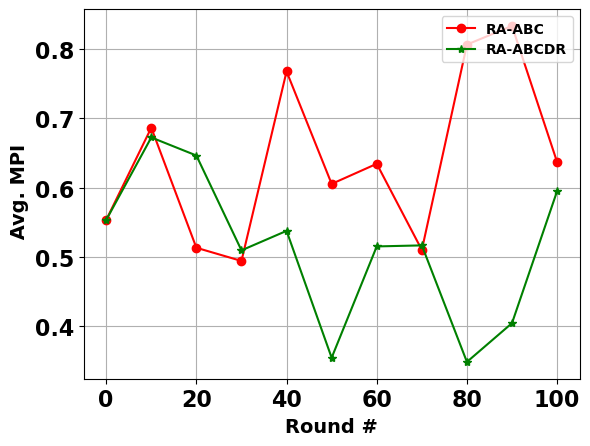}}
\caption{Monopoly Prevention Index (MPI)} 
\label{fig5b}
\end{subfigure}
\caption{System fairness analysis}\label{fig5}
\end{figure}
\hspace{0.5cm} The graph in Fig.~\ref{fig5a} compares the average Bid Accuracy Ratio (BAR) over auction rounds. Both methods start with similar BAR values, but RA-ABC steadily increases, peaking around round 60 before gradually declining, while RA-ABCDR peaks earlier (round 50) but experiences a sharp drop near round 70, stabilizing at a lower BAR by round 100. This trend suggests RA-ABC maintains greater bid accuracy stability with fewer fluctuations, whereas RA-ABCDR faces higher volatility due to dynamic recruitment effects. The sudden drop in RA-ABCDR indicates that new user integration impacts bid precision, leading to deviations in expected sensing costs. In contrast, RA-ABC's smoother trajectory implies a more consistent bidding structure, making it more suitable for long-term bid accuracy stability.
In Fig.~\ref{fig5b} the graph compares the average Monopoly Prevention Index (MPI). RA-ABC exhibits higher peaks and greater variability, suggesting fluctuations in bid fairness, while RA-ABCDR maintains more stability, ensuring a consistent distribution of auction wins. The lower volatility in RA-ABCDR implies better fairness control, minimizing monopolization risks and fostering long-term auction balance. In contrast, RA-ABC's fluctuating MPI values indicate possible fairness inconsistencies, which may impact sustained user engagement. RA-ABCDR is the better approach for ensuring equitable participation and stable bidding dynamics over multiple rounds.

\section*{Conclusion and Future Directions}\label{sec6}
\hspace{0.5cm} This paper explores incentive distribution in an MCS system using reverse auction, where participants bid, and only winners perform sensing tasks based on predefined quotas. RA-ABC allows bidding before sensing, unlike traditional systems where true values are known beforehand. RA-ABCDR enhances the system by enabling new users to join in later auction rounds, ensuring sustained engagement.
RA-ABCDR outperforms Tullock in both participant retention and auction cost efficiency. By round 100, RA-ABCDR retains 54.6\% more users than Tullock \cite{b21}, demonstrating significantly better engagement sustainability. Additionally, RA-ABCDR reduces auction costs by approximately 22.2\% compared to Tullock, ensuring greater cost-effectiveness while maintaining competitive bidding dynamics. 
RA-ABCDR improves system stability, fairness, and efficiency compared to RA-ABC by preventing participant drop-off through dynamic recruitment, ensuring consistent engagement across auction rounds. It maintains fair competition, reduces monopolization risks, and optimizes bid adjustments for cost efficiency. Unlike RA-ABC, which struggles with higher attrition and fairness fluctuations, RA-ABCDR fosters long-term balance, making it the superior choice for scalable and equitable MCS auctions.

Task complexity plays a significant role in bid cost variability and performance of the proposed models, as more complex tasks require greater effort, leading to increased uncertainty in cost estimation. Quality assurance is another crucial factor in performance, as tasks requiring high precision or multi-modal data fusion influence auction decisions. If bid costs fail to reflect actual effort, suboptimal sensing quality could impact overall system reliability. To address this, differentiated incentives can be introduced in future, allowing RA-ABC to implement weighted task pricing—ensuring users with specialized expertise or superior sensors receive appropriate compensation. Additionally, heterogeneous sensing devices introduce variability in accuracy and energy consumption, where high-fidelity sensors consume more energy but provide better data quality. Device-aware cost normalization may be necessary to prevent bias, ensuring fairness across diverse hardware. Future work can implement truncated normal distribution models to standardize bid expectations for different device capabilities to acheive this. Dynamic reputation weighting can be used to maintain auction integrity by prioritizing devices with historically reliable data, preventing low-quality sensors from winning bids solely due to lower cost estimates. By integrating these considerations, RA-ABC can enhance fairness while optimizing sensing efficiency across diverse participants. 

Our current efforts are focused to develop better processes for participants selection, improving the platform utility and considering the real world situation of risk-averse bidders in the auction system. A system for selection of different satisfactory threshold for ROI value for individuals depending on their own true value would also be more useful for improving participant utility, which is a challenge since the platform will need to know more information about each user’s device capacities which can be seen as privacy issue. 









\begin{thebibliography}{00}
\bibitem{b1}Lane, Nicholas D., Emiliano Miluzzo, Hong Lu, Daniel Peebles, Tanzeem Choudhury, and Andrew T. Campbell. "A survey of mobile phone sensing." IEEE Communications magazine 48, no. 9 (2010): 140-150.
\bibitem{b2}J. Burke et al.,“Participatory Sensing,” Wksp. World-Sensor-Web, Collocated with ACM SenSys, 2006; http://www.sensorplanet.org/wsw2006/.
\bibitem{b3}Ganti, Raghu \& Ye, Fan \& Lei, Hui. (2011). Mobile Crowd Sensing: Current State and Future Challenges. IEEE Communications Magazine. 49. 32-39. 10.1109/MCOM.2011.6069707.
\bibitem{b4}Guhya D Y. (2010), Reverse Auction Bidding: a statistical review of the first case study [Master’s Thesis]. Texas A\&M University.
\bibitem{b5}Guo B, Wang Z, Yu Z, Wang Y, Yen NY, Huang R, Zhou X (2015) Mobile crowd sensing and computing: The review of an emerging human-powered sensing paradigm. ACM Computing Surveys (CSUR) 48(1):7
\bibitem{b6}X. Zhang, Z. Yang, W. Sun, Y. Liu, S. Tang, K. Xing, and X. Mao, “Incentives for mobile crowd sensing: A survey,” IEEE Communications Surveys Tutorials 18, 54–67 (2016).
\bibitem{b8}Y. Wen, J. Shi, Q. Zhang, X. Tian, Z. Huang, H. Yu, Y. Cheng, and X. Shen, “Quality-driven auction-based incentive mechanism for mobile crowd sensing,” IEEE Transactions on Vehicular Technology 64, 4203–4214 (2015). 
\bibitem{b10}M. Xiao, B. An, J. Wang, G. Gao, S. Zhang, and J. Wu, “Cmab-based reverse auction for unknown worker recruitment in mobile crowdsensing,” IEEE Transactions on Mobile Computing, 1–1 (2021). 
\bibitem{b11}J. Liu, S. Huang, W. Wang, D. Li, and X. Deng, “An incentive mechanism based on endowment effect facing social welfare in crowdsensing,” Peer-to-Peer Networking and Applications 14, 1–17 (2021). 
\bibitem{b12}Y. Zhong and X. Zhang, “Bilateral privacy-preserving truthful incentive for mobile crowdsensing,” IEEE Systems Journal, 1–12 (2021). 
\bibitem{b13}H. Wang, Y. Yang, E. Wang, L. Wang, Q. Li, and Z. Yu, “Incentive mechanism for mobile devices in dynamic crowd sensing system,” IEEE Transactions on Human-Machine Systems 51, 365–375 (2021). 
\bibitem{b14}Y. Zhang, X. Zhang, and F. Li, “Bicrowd: Online biobjective incentive mechanism for mobile crowdsensing,” IEEE Internet of Things Journal 7, 11078–11091 (2020). 
\bibitem{b15} J. -S. Lee and B. Hoh, "Sell your experiences: a market mechanism based incentive for participatory sensing," 2010 IEEE International Conference on Pervasive Computing and Communications (PerCom), Mannheim, Germany, 2010, pp. 60-68, doi: 10.1109/PERCOM.2010.5466993.
\bibitem{b16} E. Wang, H. Wang, Y. Yang and W. Liu, "Truthful Incentive Mechanism for Budget-Constrained Online User Selection in Mobile Crowdsensing," in IEEE Transactions on Mobile Computing, vol. 21, no. 12, pp. 4642-4655, 1 Dec. 2022, doi: 10.1109/TMC.2021.3083920.
\bibitem{b17}Perry, Marcus. (2010). The Exponentially Weighted Moving Average. 10.1002/9780470400531.eorms0314. 
\bibitem{b18} R. McAfee and PJ. McMillan, "Auction and Bidding", 1. Economic Literature, 25: 699 - 738, 1997.
\bibitem{b19} Sasank Reddy, Deborah Estrin, Mark Hansen, and Mani Srivastava. 2010. Examining micro-payments for participatory sensing data collections. In Proceedings of the 12th ACM international conference on Ubiquitous computing (UbiComp '10). Association for Computing Machinery, New York, NY, USA, 33–36. https://doi.org/10.1145/1864349.1864355
\bibitem{b20} Zamfir, Mariana and Manea, Marinela and Ionescu, Luiza. (2016). Return On Investment – Indicator for Measuring the Profitability of Invested Capital. Valahian Journal of Economic Sciences. 7. 10.1515/vjes-2016-0010. 
\bibitem{b21} Luo, T., Kanhere, S. S., Tan, H.-P., Wu, F., \& Wu, H. (2015). Crowdsourcing with Tullock contests: A new perspective. Proceedings of the IEEE Conference on Computer Communications (INFOCOM), 2515–2525. https://doi.org/10.1109/INFOCOM.2015.7218555
\bibitem{b22} U.S. Department of Justice \& Federal Trade Commission. (2010). Horizontal merger guidelines.
\end{thebibliography}
\end{document}